# Phenomenological Theory of a Single Domain Wall in Uniaxial Trigonal Ferroelectrics: Lithium Niobate and Lithium Tantalate

David A Scrymgeour and Venkatraman Gopalan

Department of Materials Science and Engineering, The Pennsylvania State University

Amit Itagi, Avadh Saxena, and Pieter J Swart

Theoretical Division, Los Alamos National Laboratory

A phenomenological treatment of domain walls based on the Ginzburg-Landau-Devonshire theory is developed for uniaxial, trigonal ferroelectrics lithium niobate and lithium tantalate. The contributions to the domain wall energy from polarization and strain as a function of orientation are considered. Analytical expressions are developed which are analyzed numerically to determine the minimum polarization, strain, and energy configurations of domain walls. It is found that hexagonal *y*-walls are preferred over *x*-walls in both materials. This agrees well with experimental observation of domain geometries in *stoichiometric* composition crystals.

# 1. Introduction

Recently, considerable attention has been focused on the phenomena of antiparallel (180°) ferroelectric domains in ferroelectrics lithium niobate (LiNbO$_3$) and lithium tantalate (LiTaO$_3$) and their manipulation into diverse shapes on various length scales.[1] For example, optical and acoustic frequency conversion devices require periodic gratings of antiparallel domains,[2] and electro-optic devices require domains to be shaped as lenses and prisms.[3] Therefore, the structure of a domain walls in these materials has become an important subject of study.[4,5] The hexagonal unit cell and the atomic arrangement in the basal plane are shown in Figure 1.

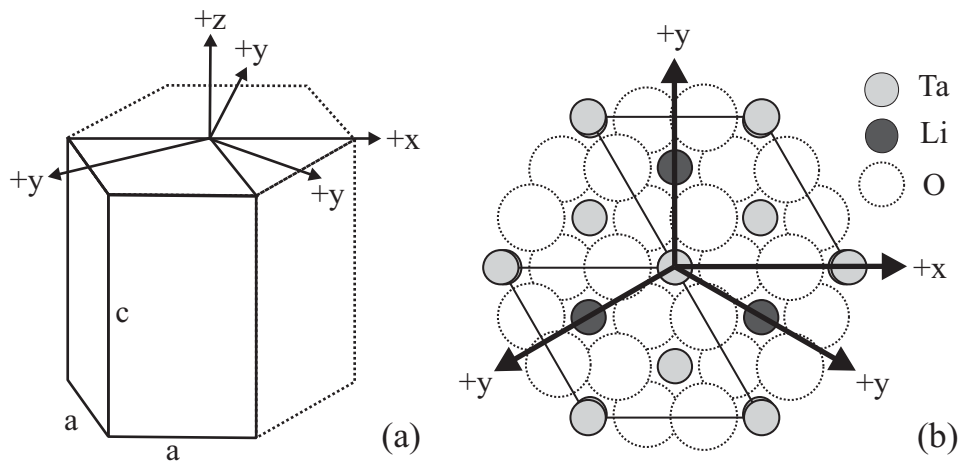

**Figure 1: (a) A schematic of the hexagonal unit cell of ferroelectric LiTaO$_3$ (space group R3c) where *a* and *c* are the lattice parameters in the hexagonal notation. (b) The arrangement of the atoms projected on the (0001) plane, where a solid trapezoid is a unit cell.**

From a fundamental viewpoint, the domain wall structure and shapes observed in these materials highlight interesting issues relating to preferred domain wall orientations, wall strains, wall width, and defect mediated changes in the local structure of these domain walls. For example, when domains are created at room temperature in a single crystal of $LiNbO_3$ or $LiTaO_3$ by external electric fields, one observes a variety of naturally preferred crystallographic shapes exhibited by these crystals depending on slight variations in crystal stoichiometry. These single crystals are typically either of congruent composition which are deficient in lithium (Li/(Li+Nb, Ta) ~ 0.485) or of stoichiometric composition ( Li/(Li+Nb, Ta)=0.5). Figure 2 summarizes the temperature and stoichiometry dependence of the two primary shapes preferred by these crystals. The *stoichiometric* crystals of both $LiNbO_3$ and $LiTaO_3$ exhibit six-sided polygonal shapes, with domain walls parallel to the crystallographic *c*-glide planes (*yz*-plane), termed as *y*-walls (as pictured in Figure 2(a)). With lithium deficiency in the crystals, the shape of the domains in *congruent* $LiTaO_3$ changes to triangular domains with domain walls parallel to the crystallographic *xz*-planes, termed *x*-walls as shown in Figure 2(c). This change in domain shape is not seen in *congruent* $LiNbO_3$. When the domains are created at higher temperatures (>125ºC), the congruent $LiTaO_3$ crystals form hexagonal domains of *y*-wall orientation, the same as the wall orientations in congruent and stoichiometric $LiNbO_3$. It is also important to note that with increasing lithium deficiency in the crystals, the regions adjoining domain walls show increased optical birefringence,[6] strains,[7] and local electric fields that extend over microns. These phenomena have been shown to arise from aggregated point defect complexes in the material that transition from a frustrated to a stable defect state across a wall at room temperature.[4] At higher temperatures (>100ºC), these defect complexes break up and the domain wall strains, optical birefringence, and local fields disappear as well.

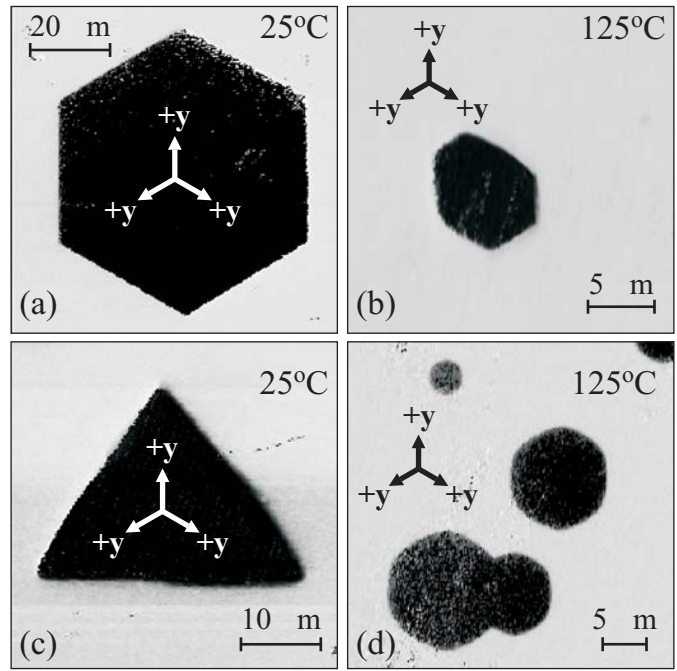

**Figure 2: Piezoelectric force microscopy[8] phase contrast images of domain shapes created in (a),(b) congruent LiNbO$_3$ and (c),(d) congruent LiTaO$_3$. Domains in (a) and (c) created at room temperature and (b) and (d) created at 125°C.**

The above observations are driven by both crystallographic considerations and defect-mediated changes. Towards separating these effects, this paper addresses the following question: what are the energetically favored orientations of domain walls in *stoichiometric* LiNbO$_3$ and LiTaO$_3$ purely from a crystallographic viewpoint. We will assume that there is no external electric field applied and the crystal is unclamped. Both LiTaO$_3$ and LiNbO$_3$ show a second order phase transition from a higher temperature paraelectric phase with space group symmetry $R\bar{3}c$ to a ferroelectric phase of symmetry

R3c at Curie temperatures $T_c$ of ~ 690° C and ~1190°C, respectively. The approach is to minimize the invariant Ginzburg-Landau-Devonshire (GLD) free energy for a crystal in the presence of a single 180° domain wall. This yields the strains, wall width, and the minimum energy orientations of this wall, which can then be compared with the experimental observations. General conclusions can also be drawn regarding the possible reasons for domain shape changes introduced by the addition of defects.

The outline of the paper is as follows. The theoretical framework for the analysis is presented in Section 2. The equilibrium values of the polarization and the strain fields in the case of a homogeneous sample without any domain walls are derived in Section 2.2. In Section 2.3, a single domain wall is introduced in the sample and the nature of the polarization and strain fields in the domain wall is derived. The numerical results are presented in Section 3. These results are discussed in Section 4.

## 2. Theoretical Framework

We base our analysis on the Ginzburg-Landau-Devonshire (GLD) theory.[9,10] According to the Landau theory, the phase transition from the paraelectric phase to the ferroelectric phase occurs as a result of symmetry breaking. In LiTaO$_3$ and LiNbO$_3$, the paraelectric phase belongs to the space group $R\bar{3}c$ $\left(D_{3d}^6\right)$ and the ferroelectric phase belongs to the space group $R3c$ $\left(C_{3v}^6\right)$ (loss of inversion symmetry). The symmetry breaking results in the evolution of a primary order parameter in the low symmetry ferroelectric phase. In the case of LiTaO$_3$ and LiNbO$_3$, the primary order parameter is the polarization along the crystallographic z direction, $P_z$. This order parameter transforms as the basis function of the $\Gamma_2^-$ irreducible representation of the prototype phase space group and the other two components ($P_x$, $P_y$) belong to the $\Gamma_3^-$ irreducible representation. The area change (compression or dilatation) of the hexagonal basal plane and the elaongation along the z-axis both belong to a one-dimensional irreducible representation of strain $\Gamma_1^+$. Similarly, the two shears of the basal plane and the two shears in the *x-y* and *y-z* plane both belong to a two-dimensional irreducible representation of strain $\Gamma_3^+$.

The fields of interest are the macroscopic strains and the macroscopic polarization. The six strain components and the two orthogonal components of the polarization other than the primary order parameter are coupled to the primary order parameter and are treated as secondary order parameters in our analysis.

Since we are interested in the macroscopic fields, we are only interested in the $\Gamma$ point (zone center) in the Brillouin zone. Thus, the symmetry considerations for the free energy reduce to the considerations of the symmetry of the point group of the prototype phase $\bar{3}m$ (D$_{3d}$). The presence of domain walls can be considered as perturbations in the vicinity of the $\Gamma$ point. This is reflected in the free energy that corresponds to the gradient in the order parameters in the GLD theory.

The approach adopted here is as follows. We first determine the free energy that must be invariant under the prototype point group symmetry operations. We minimize this free energy with the polarization components as variables with the constraint that the crystal is stress-free. This gives the equilibrium values of polarization and strain. Using the homogeneous values of the polarization and the strain components, we then introduce an infinite 180° domain wall at some angle to the crystallographic *x-z* plane. The structure of the domain wall is obtained using variational minimization of the total free energy under the constraints of strain compatibility and mechanical equilibrium.

## 2.1 Free Energy

The general form of the free energy of a ferroelectric material is given by the equation

$$F(P_i, P_{i,j}, \varepsilon_k) = F_L(P_i) + F_{el}(\varepsilon_k) + F_c(P_i, \varepsilon_k) + F_G(P_{i,j}). \tag{1}$$

where $P_i$ are the polarization components, and $\varepsilon_k$ are the strains in Voigt's notation. In particular, LiNbO$_3$ and LiTaO$_3$ belong to the $\bar{3}m$ point group. In the following analysis, the crystallographic uniaxial direction is denoted as $z$-axis. The $y$-axis is chosen such that the $y$-$z$ plane coincides with a crystal-glide plane as shown in Figure 1. The $x$-axis is chosen such that the $x$, $y$ and $z$ axes form a right-handed Cartesian coordinate system. The free energy form that is invariant under the point group $\bar{3}m$ consists of the following terms: The first term is the Landau-Devonshire free energy describing a second order phase transition,[9] and is given by

$$F_L(P_i) = -\frac{\alpha_1}{2} P_z^2 + \frac{\alpha_2}{4} P_z^4 + \frac{\alpha_3}{2}(P_x^2 + P_y^2), \tag{2}$$

where $\alpha_1$ is temperature dependent and positive in the ferroelectric phase, while $\alpha_2$ and $\alpha_3$ are positive. The $\alpha_i$ are given in Table 1 which are related to the dielectric constants given in Table 2. The elastic free energy of the system is given by

$$\begin{aligned}F_{el}(\varepsilon_k) &= \beta_1 \varepsilon_3^2 + \beta_2(\varepsilon_1 + \varepsilon_2)^2 + \beta_3\left[(\varepsilon_1 - \varepsilon_2)^2 + \varepsilon_6^2\right] \\ &+ \beta_4 \varepsilon_3(\varepsilon_1 + \varepsilon_2) + \beta_5(\varepsilon_4^2 + \varepsilon_5^2) + \beta_6\left[(\varepsilon_1 - \varepsilon_2)\varepsilon_4 + \varepsilon_5 \varepsilon_6\right]\end{aligned}, \tag{3}$$

where, following Voigt's notation, $\varepsilon_1 = u_{1,1}$, $\varepsilon_2 = u_{2,2}$, $\varepsilon_3 = u_{3,3}$, $\varepsilon_4 = u_{2,3} + u_{3,2}$, $\varepsilon_5 = u_{1,3} + u_{3,1}$, and $\varepsilon_6 = u_{1,2} + u_{2,1}$, and $u_i$ are the lattice displacements. The $\beta_i$ are given in Table 1 related to the elastic constants given in Table 2. The third term in Eq. (1) is the electrostrictive coupling between the polarization and strain components and is given by

$$F_c(P_i, \varepsilon_k) = \gamma_1(\varepsilon_1 + \varepsilon_2)P_z^2 + \gamma_2\varepsilon_3 P_z^2 + \gamma_3\left[(\varepsilon_1 - \varepsilon_2)P_y P_z + \varepsilon_6 P_x P_z\right] + \gamma_4(\varepsilon_5 P_x P_z + \varepsilon_4 P_y P_z)$$

$$+ \gamma_5(\varepsilon_1 + \varepsilon_2)(P_x^2 + P_y^2) + \gamma_6\varepsilon_3(P_x^2 + P_y^2) + \gamma_7\left[(\varepsilon_1 - \varepsilon_2)(P_x^2 - P_y^2) + 2\varepsilon_6 P_x P_y\right]$$

$$+ \gamma_8\left[\varepsilon_4(P_x^2 - P_y^2) + 2\varepsilon_5 P_x P_y\right] \tag{4}$$

where the $\gamma_i$ are listed in Table 1 which are related to the electrostrictive and elastic constants given in Table 2. The final term in Eq. (1) is the gradient energy of the lowest order compatible with the $\overline{3}m$ symmetry, and is given by

$$F_G(P_{i,j}) = g_1(P_{z,1}^2 + P_{z,2}^2) + g_2(P_{z,3}^2) \tag{5}$$

Here, $g_1$ and $g_2$ are the gradient coefficients. To keep the mathematical complexity tractable at this stage, we neglect the energy contribution from the gradient of the secondary order parameters. We will neglect the electrostrictive coupling energy terms from Eq. (4) that do not involve the primary order parameter, $P_z$. Later in Section 3.3 we show that gradient terms of the type $P_{n,n}$ play an important role in determining the domain shape as well. The gradient term captures short range interactions. However, while considering an inhomogeneous case, non-local (or long-range) electric dipole-dipole interaction must be included in principle.[11] Not including this interaction slightly changes the profile and energetics of the domain wall.

In the presence of a domain wall at a variable orientation to the *x*-axis or *y*-axis, it is convenient to work in a rotated coordinate system as shown in Figure 3. This new coordinate system is obtained by a proper rotation of the *x*-axis and *y*-axis about the *z*-axis, such that $x \to x_n$ and $y \to x_t$ and $(x_n, x_t, z)$ forms a right handed coordinate system. The subscripts *n* and *t*, respectively, refer to the coordinates normal and parallel to the domain wall. The free energy in the new coordinate system is then given by

$$F(P_i, \varepsilon_k, P_{i,j}) = -\frac{\alpha_1}{2} P_z^2 + \frac{\alpha_2}{4} P_z^4 + \frac{\alpha_3}{2}(P_n^2 + P_t^2)$$

$$+ \beta_1 \varepsilon_3^2 + \beta_2(\varepsilon_n + \varepsilon_t)^2 + \beta_3\left[(\varepsilon_n - \varepsilon_t)^2 + \tilde{\varepsilon}_6^2\right] + \beta_4 \varepsilon_3(\varepsilon_n + \varepsilon_t)$$

$$+ \beta_5(\tilde{\varepsilon}_4^2 + \tilde{\varepsilon}_5^2) + \beta_6\left[(\varepsilon_n - \varepsilon_t)\tilde{\varepsilon}_4 + \tilde{\varepsilon}_5\tilde{\varepsilon}_6\right]\cos(3\theta) + \beta_6\left[(\varepsilon_n - \varepsilon_t)\tilde{\varepsilon}_5 - \tilde{\varepsilon}_4\tilde{\varepsilon}_6\right]\sin(3\theta)$$

$$+ \gamma_1(\varepsilon_n + \varepsilon_t)P_z^2 + \gamma_2 \varepsilon_3 P_z^2 + \gamma_3\left[(\varepsilon_n - \varepsilon_t)P_t P_z + \tilde{\varepsilon}_6 P_n P_z\right]\cos(3\theta)$$

$$+ \gamma_3\left[(\varepsilon_n - \varepsilon_t)P_n P_z - \tilde{\varepsilon}_6 P_t P_z\right]\sin(3\theta) + \gamma_4(\tilde{\varepsilon}_5 P_n P_z + \tilde{\varepsilon}_4 P_t P_z)$$

$$+ g_1(P_{z,n}^2 + P_{z,t}^2) + g_2(P_{z,3}^2) \qquad (6)$$

where $\theta$ is the angle between the *x* and $x_n$ coordinate axes. Following Voigt's notation, $\varepsilon_n = u_{n,n}$, $\varepsilon_t = u_{t,t}$, $\varepsilon_3 = u_{3,3}$, $\tilde{\varepsilon}_4 = u_{t,3} + u_{3,t}$, $\tilde{\varepsilon}_5 = u_{n,3} + u_{3,n}$, and $\tilde{\varepsilon}_6 = u_{t,n} + u_{n,t}$, $u_i$ are the lattice displacements, and $P_n$ and $P_t$ are polarizations along the *n* and *t* axes, respectively. The following analysis will use the free energy in Eq. (6).

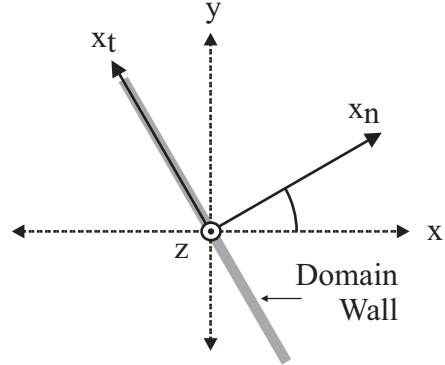

**Figure 3:** Orientation of the rotated coordinate system ($x_n$, $x_t$, $z$) with respect to the crystallographic coordinate system ($x,y,z$). Also noted is the domain wall orientation, which is parallel to the $x_t$ axis.

**Table 1: Derived Constants in Eqs. (2)-(4).**

| | Expression | LiTaO$_3$ | LiNbO$_3$ | Units |
|---|---|---|---|---|
| $\alpha_1$ | $1/2\varepsilon_{33}$ | 1.256 | 2.012 | $\times 10^9$ Nm$^2$/C$^2$ |
| $\alpha_2$ | * derived from Equation 11 | 5.043 | 3.608 | $\times 10^9$ Nm$^6$/C$^4$ |
| $\alpha_3$ | $1/\varepsilon_{11}$ | 2.22 | 1.345 | $\times 10^9$ Nm$^2$/C$^2$ |
| $\beta_1$ | $\frac{1}{2}C_{33}$ | 13.55 | 12.25 | $\times 10^{10}$ N/m$^2$ |
| $\beta_2$ | $\frac{1}{4}(C_{11}+C_{12})$ | 6.475 | 6.4 | $\times 10^{10}$ N/m$^2$ |
| $\beta_3$ | $\frac{1}{4}(C_{11}-C_{12})$ | 4.925 | 3.75 | $\times 10^{10}$ N/m$^2$ |
| $\beta_4$ | $C_{13}$ | 7.4 | 7.5 | $\times 10^{10}$ N/m$^2$ |
| $\beta_5$ | $\frac{1}{2}C_{44}$ | 4.8 | 3 | $\times 10^{10}$ N/m$^2$ |
| $\beta_6$ | $C_{14}$ | -1.2 | 0.9 | $\times 10^{10}$ N/m$^2$ |
| $\gamma_1$ | $\frac{1}{2}(C_{11}+C_{12})Q_{31}+\frac{1}{2}C_{13}Q_{33}$ | -0.202 | 0.216 | $\times 10^9$ N m$^2$/C$^2$ |
| $\gamma_2$ | $\frac{1}{2}C_{33}Q_{33}+\frac{1}{2}C_{13}Q_{31}$ | 1.317 | 1.848 | $\times 10^9$ N m$^2$/C$^2$ |
| $\gamma_3$ | $2C_{14}Q_{44}-\frac{1}{2}(C_{11}-C_{12})Q_{42}$ | -2.824 | -0.33 | $\times 10^9$ N m$^2$/C$^2$ |
| $\gamma_4$ | $C_{44}Q_{44}$ | 4.992 | 3.9 | $\times 10^9$ N m$^2$/C$^2$ |
| $\lambda_1$ | - | 6.418 | 9.359 | $\times 10^{-4}$ |
| $\lambda_2$ | - | -0.157 | -0.4874 | $\times 10^{-4}$ |

**Table 2: Relevant physical constants of LiNbO$_3$ and LiTaO$_3$**

|  | LiTaO$_3$ [12,13] | LiNbO$_3$ [12,14] | Units |
|---|---|---|---|
| P$_s$ | 50-55 | 70-75 | μC/cm$^2$ |
| ε$_{11}$ | 52.7 ± 1.1 | 84.3 ± 0.8 | - |
| ε$_{33}$ | 44.0 ± 0.7 | 28.9 ± 0.7 | - |
| C$_{11}$ | 2.3305 ± 0.0004 | 1.9886 ± 0.0003 | ×10$^{11}$ N/m$^2$ |
| C$_{12}$ | 0.4644 ± 0.0006 | 0.5467 ± 0.0004 | ×10$^{11}$ N/m$^2$ |
| C$_{13}$ | 0.8358 ± 0.0063 | 0.6726 ± 0.0093 | ×10$^{11}$ N/m$^2$ |
| C$_{33}$ | -2.7414 ± 0.0104 | 2.3370 ± 0.0152 | ×10$^{11}$ N/m$^2$ |
| C$_{14}$ | -1.067 ± 0.0004 | 0.0783 ± 0.0002 | ×10$^{11}$ N/m$^2$ |
| C$_{44}$ | 0.9526 ± 0.0002 | 0.5985 ± 0.0001 | ×10$^{11}$ N/m$^2$ |
| Q$_{31}$ | -0.00485 ± 0.0002 | -0.003 | m$^4$/C$^2$ |
| Q$_{33}$ | 0.016 ± 0.007 | 0.016 | m$^4$/C$^2$ |
| Q$_{42}$ | 0.016 ± 0.0001 | -0.003 ± 0.03 | m$^4$/C$^2$ |
| Q$_{44}$ | 0.056 ± 0.005 | 0.0375 ± 0.03 | m$^4$/C$^2$ |

**2.2. Homogeneous case: Single domain state**

We first consider the homogeneous case where the material exists in a single domain state and apply the following constraints

$$\frac{\partial F}{\partial P_i} = 0, \tag{7}$$

$$\frac{\partial F}{\partial \varepsilon_i} = \sigma_i = 0 \tag{8}$$

where $\sigma_i$ is the stress. Constraint (7) specifies uniform homogeneous polarization values in the material and (8) specifies that the material is stress free. These constraints result in the following homogeneous strains and polarizations:

$$\tilde{\varepsilon}_4 = \tilde{\varepsilon}_5 = \tilde{\varepsilon}_6 = 0, \tag{9}$$

$$P_n = P_t = 0, \tag{10}$$

$$P_z = P_h = \pm \left[ \frac{\alpha_1}{\alpha_2 + 4(\beta_1 \psi_2^2 + 4\beta_2 \psi_1^2 + 2\beta_4 \psi_1 \psi_2 + 2\gamma_1 \psi_1 + \gamma_2 \psi_2)} \right]^{1/2}, \tag{11}$$

where the subscript $h$ refers to the homogeneous case and

$$\psi_1 = \frac{2\gamma_1 \beta_1 - \gamma_2 \beta_4}{2(\beta_4^2 - 4\beta_1 \beta_2)} \quad \text{and} \quad \psi_2 = \frac{2\gamma_2 \beta_2 - \gamma_1 \beta_4}{\beta_4^2 - 4\beta_1 \beta_2} \tag{12}$$

Using the homogeneous value $P_z = P_h$ of $z$-axis polarization, we obtain the spontaneous dilatory strains as

$$\varepsilon_n = \varepsilon_t = \lambda_1 = \psi_1 P_h^2 \tag{13}$$

$$\varepsilon_3 = \lambda_2 = \psi_2 P_h^2 \tag{14}$$

It can be seen that in the homogeneous case, there is no polarization in the *n-t* plane and that the shear strains are zero. There are two possible orientations for the homogeneous polarization, $P_h$. Note that $P_h$ is equal to the spontaneous polarization value, $P_s$, as found in the literature.[13,14] The coefficient $\alpha_2$ in Table 1 was determined using Eq. (11) and the known experimental values of $\alpha_1$ and $P_h$ at room temperature for LiNbO$_3$ and LiTaO$_3$. The values of $\lambda_1$, and $\lambda_2$ are, respectively, $\lambda_1 = 6.4 \times 10^{-4}$ and $\lambda_2 = -1.6 \times 10^{-3}$ (for LiTaO$_3$) and $\lambda_1 = 9.36 \times 10^{-4}$ and $\lambda_2 = -4.8 \times 10^{-3}$ (for LiNbO$_3$), indicating that there is a homogeneous tensile strain in the *x-y* plane and a homogeneous compressive strain in the *z*-direction.

### 2.3. Inhomogeneous Case: A single infinite domain wall

We now introduce an infinite 180° domain wall in the crystal. The position of the wall in the rotated coordinate system is shown in Figure 3. The $x_t$-$z$ plane corresponds to the plane in the domain wall where the *z*-component of the polarization vanishes. Far away from the domain wall, we assume that the polarizations take a homogeneous value of $-P_h$ in the $-x_n$ direction and $+P_h$ in the $+x_n$ direction. The angle $\theta$ between the normal to the domain wall, $x_n$ with the crystallographic *x-axis* defines the orientation of the wall in the *x-y* plane. We will seek a quasi-one dimensional solution, where the polarization and strain fields are functions of only the coordinate normal to the wall (i.e. the coordinate $x_n$). In a defect free material, the St. Venant's strain compatibility condition must hold

$$\nabla \times \nabla \times \overleftrightarrow{\varepsilon} = 0, \qquad (15)$$

where $\overleftrightarrow{\varepsilon}$ in the above equation is the strain tensor.[15] Noting that the strains are a function of $x_n$ only, and taking the homogeneous values far away from the wall, Eq. (15) yields

$$\tilde{\varepsilon}_4 = 0, \ \varepsilon_t = \lambda_1, \ \varepsilon_3 = \lambda_2 \tag{16}$$

Note that these strain values are valid throughout the material including the wall region. In addition, the divergence of stress must be zero to ensure mechanical equilibrium, i.e.

$$\nabla \cdot \overleftrightarrow{\sigma} = 0, \tag{17}$$

where $\overleftrightarrow{\sigma}$ represents the stress tensor. Noting that the stresses are a function of $x_n$ only, and vanish far away from the wall, Eq. (17) yields

$$\sigma_n = \tilde{\sigma}_5 = \tilde{\sigma}_6 = 0 \tag{18}$$

Defining $\Delta\varepsilon_n = \varepsilon_n - \lambda_1$, as the deviation of the normal strain $\varepsilon_n$ from the homogeneous value $\lambda_1$, Eq. (18) gives,

$$\begin{bmatrix} \Delta\varepsilon_n \\ \tilde{\varepsilon}_5 \\ \tilde{\varepsilon}_6 \end{bmatrix} = [m_{ij}] \begin{bmatrix} P_z^2 - P_h^2 \\ P_z P_n \\ P_z P_t \end{bmatrix} \tag{19}$$

The strains $\tilde{\varepsilon}_5$ and $\tilde{\varepsilon}_6$ can also be considered as deviations from their homogeneous values, recalling that their homogeneous values are zero from Eq. (9). The matrix $[m_{ij}] = [a_{ij}]^{-1}[b_{ij}]$, where

$$[a_{ij}] = \begin{bmatrix} 2(\beta_2 + \beta_3) & \beta_6 \sin(3\theta) & 0 \\ \beta_6 \sin(3\theta) & 2\beta_5 & \beta_6 \cos(3\theta) \\ 0 & \beta_6 \cos(3\theta) & 2\beta_3 \end{bmatrix}, \tag{20}$$

$$[b_{ij}] = \begin{bmatrix} -\gamma_1 & -\gamma_3 \sin(3\theta) & -\gamma_3 \cos(3\theta) \\ 0 & -\gamma_4 & 0 \\ 0 & -\gamma_3 \cos(3\theta) & \gamma_3 \sin(3\theta) \end{bmatrix}. \quad (21)$$

Now we minimize the total free energy F with respect to the polarizations $P_n$ and $P_t$ as follows:

$$\frac{\partial F}{\partial P_i} = 0, \ (i=n,t) \quad (22)$$

where, for the present, the gradient terms $P_{n,n}$, $P_{n,t}$, $P_{t,n}$, and $P_{t,t}$ have been ignored. Equation (22) in combination with Eqs. (16) and Eqs. (19)-(21) yields relationships between the polarizations $P_n$, $P_t$, and $P_z$ as follows:

$$\begin{bmatrix} P_n \\ P_t \end{bmatrix} = \frac{P_z(P_z^2 - P_h^2)}{\alpha_3^2 + \alpha_3(\mu_{11} + \mu_{22})P_z^2 + (\mu_{11}\mu_{22} - \mu_{12}\mu_{21})P_z^4} \begin{bmatrix} v_1\alpha_3 + (v_1\mu_{22} - v_2\mu_{12})P_z^2 \\ v_2\alpha_3 + (v_2\mu_{11} + v_1\mu_{21})P_z^2 \end{bmatrix}. \quad (23)$$

The constants $v_i$ and $\mu_{ij}$ are listed in Appendix A.

From Eq. (23), we see that the polarizations $P_n$ and $P_t$ depend on $P_z$ in a highly nonlinear manner. In order to simplify these relations for further progress, we estimate the relative magnitudes of different terms in the denominator of the prefactor in Eq. (23) for $0 \leq \theta \leq 2\pi$ and $|P_z| \leq P_h$. Using the values of physical constants given in Tables 1 and 2 for LiNbO$_3$ and LiTaO$_3$, we find that $\alpha_3^2 \sim 10^{19}$ N$^2$m$^4$C$^{-4}$, $|\alpha_3(\mu_{11} + \mu_{22})P_z^2| \leq 10^{18}$ N$^2$m$^4$C$^{-4}$, and $|(\mu_{11}\mu_{22} - \mu_{12}\mu_{21})P_z^4| \leq 10^{16}$ N$^2$m$^4$C$^{-4}$. Therefore, we retain only the $\alpha_3^2$ term in the denominator of the prefactor in Eq. (23). The polarizations $P_n$ and $P_t$ simplify to odd functions of $P_z$ and vanish at $P_z=P_h$. From Eq. (19), we note that the strains $\Delta\varepsilon_n$, $\tilde{\varepsilon}_5$, and $\tilde{\varepsilon}_6$ are even functions of $P_z$ and vanish at $P_z=P_h$.

$$\begin{bmatrix} P_n \\ P_t \end{bmatrix} = [\rho_{ij}] \begin{bmatrix} P_z \\ P_z^3 \\ P_z^5 \end{bmatrix}, \tag{24}$$

$$\begin{bmatrix} \Delta\varepsilon_n \\ \tilde{\varepsilon}_5 \\ \tilde{\varepsilon}_6 \end{bmatrix} = [\phi_{ij}] \begin{bmatrix} 1 \\ P_z^2 \\ P_z^4 \\ P_z^6 \end{bmatrix} \tag{25}$$

where the matrices $[\rho_{ij}]$ and $[\phi_{ij}]$ are listed in the Appendix A.

So far, we have minimized the total free energy, $F$, with respect to $P_n$ and $P_t$ (Eq. (22)) under the constraints of strain compatibility (Eq. (15)) and mechanical equilibrium (Eq. (17)). This has enabled us to obtain the expressions for the secondary order parameters ($P_n$, $P_t$, and $\varepsilon_i$) in terms of the primary order parameter, $P_z$. We now perform variational minimization of the total free energy, $F$, with respect to the primary order parameter, $P_z$ under the boundary conditions that $P_z$ approaches $\pm P_h$ far away from the domain wall. This gives us the Euler-Lagrange equation,

$$\frac{\partial F}{\partial P_z} - \frac{\partial}{\partial x_n}\left(\frac{\partial F}{\partial P_{z,n}}\right) = 0. \tag{26}$$

The partial derivative $\partial F/\partial P_z$ is a polynomial in odd powers of $P_z$ as follows:

$$\frac{\partial F}{\partial P_z} = -\varsigma_1 P_z + \varsigma_3 P_z^3 + \varsigma_5 P_z^5 + \varsigma_7 P_z^7 + \varsigma_9 P_z^9 + \varsigma_{11} P_z^{11} \tag{27}$$

The first two coefficients $\varsigma_1$ and $\varsigma_3$ are given by

$$\varsigma_1 = \alpha_1 - 4\gamma_1\lambda_1 - 2\gamma_1\phi_{11} - 2\gamma_2\lambda_2 - \gamma_3(\phi_{11}\rho_{21} + \phi_{31}\rho_{11})\cos(3\theta)$$

$$- \gamma_3(\phi_{11}\rho_{11} - \phi_{31}\rho_{21})\sin(3\theta) - \gamma_4\phi_{21}\rho_{11}, \tag{28}$$

$$\varsigma_3 = \alpha_2 + 2\gamma_1\phi_{12} + \gamma_3(\phi_{11}\rho_{22} + \phi_{12}\rho_{21} + \phi_{31}\rho_{12} + \phi_{32}\rho_{11})\cos(3\theta) +$$
$$\gamma_3(\phi_{11}\rho_{12} + \phi_{12}\rho_{11} - \phi_{31}\rho_{22} - \phi_{32}\rho_{21})\sin(3\theta) + \gamma_4(\phi_{21}\rho_{12} + \phi_{22}\rho_{11}) \quad (29)$$

For further analysis of the order parameter, we truncate the polynomial in Eq. (27) after the $P_z^3$ term. On substituting for the physical properties of LiNbO$_3$ and LiTaO$_3$ from Table 1, it is found that for all values of $0 \leq \theta \leq 2\pi$, $|\varsigma_1| \sim 10^9$ Nm$^2$C$^{-2}$, $|\varsigma_3 P_h^2| \sim 10^9$ Nm$^2$C$^{-2}$, $|\varsigma_5 P_h^4| \sim 10^3$ Nm$^2$C$^{-2}$, $|\varsigma_7 P_h^6| \sim 10^2$ Nm$^2$C$^{-2}$, $|\varsigma_9 P_h^8| \sim 0.1-1$ Nm$^2$C$^{-2}$, $|\varsigma_{11} P_h^{10}| \sim 10^{-3} - 10^{-2}$ Nm$^2$C$^{-2}$. Therefore the truncation of Eq. (27) is justified. With this truncation, Eq. (27) can be rewritten as

$$2g_1 P_{z,n} = -\varsigma_1 P_z + \varsigma_3 P_z^3 \quad (30)$$

The solution to this equation is the kink, given by

$$P_z(x_n) = \sqrt{\frac{\varsigma_1}{\varsigma_3}} \tanh\left(\frac{x_n}{2}\sqrt{\frac{\varsigma_1}{g_1}}\right) \quad (31a)$$

where $x_n$ is the coordinate parallel to the domain wall normal $n$. The domain wall half-width, $x_o$ is defined as $x_o = 2\sqrt{\frac{g_1}{\varsigma_1}}$. Substituting the expression for $P_z(x_n)$ into the Eqs. (24) and (25), we get the variation of strains and in-plane polarizations, $P_n$ and $P_t$ as a function of $x_n$. Substituting these expressions into (6), we get the total inhomogeneous free energy $F_{inh}$. As a cautionary note, although in deriving Eq. (31a), we neglected the higher order terms in $P_z$ in Eq. (27), one cannot do so in calculating the total free energy, $F_{inh}$. As will be seen further on, the *variation* of the free energy, $F_{DW}$ calculated from Eq. (30) as a function of the domain wall angle $\theta$ is small as compared to the mean value itself. Therefore, the truncation of higher order polarization terms in Eq. (25) should be carried out with care, if at all. The *average* domain wall energy per unit volume, $F_{DW}$, due to the addition of a domain wall to the homogeneous single domain state can then be calculated as

$$F_{DW} = \frac{1}{\Delta x} \int_{-\Delta x/2}^{+\Delta x/2} (F_{inh} - F_h) dx_n \tag{31b}$$

where $F_{inh}$ and $F_h$ are the total free energy (Eq. (6)) for the inhomogeneous and the homogeneous states, respectively. The integration window, $\Delta x$, was chosen across the wall as $\Delta x = 4x_o$, where $x_o$ is the wall half width. This window corresponds to where the energy drops to 2.2% of the peak value at the domain wall. Integration over a larger window does not significantly increase the integrated energy. We note that after performing the integration in Eq. (31b), the $F_{DW} \propto \sqrt{g_1}$, where the other gradient term $g_2$ is ignored as before.

The general solution to Eq (30) is a kink-antikink lattice (or a "polarization grating") solution

$$P_z(x_n) = \sqrt{\frac{\varsigma_1}{\varsigma_3}} \sqrt{\frac{2k^2}{1+k^2}} sn\left(\frac{x_n}{x_L}, k\right) \tag{32}$$

where $sn(x,k)$ is a Jacobi elliptic function with modulus $k$, and for peridocity $4x_L K(k)$, where $K(k)$ is the complete elliptic of the first kind.[16] Here $x_L = x_o \sqrt{\frac{1}{2}(1+k)^2}$ and $0 \le k = \frac{P_3}{P_4} \le 1$ where $P_3$ and $P_4$ are the two positive roots of the equation: $f_o = -\left(\frac{\varsigma_1}{2}\right)P_z^2 + \left(\frac{\varsigma_3}{4}\right)P_z^4$ with $\left(\frac{-\varsigma_1^2}{4\varsigma_3}\right) \le f_o \le 0$. In the limit $k \to 1$ we recover the single kink solution of Eq. 31(a). The domain lattice energy per period can be calculated using equation 31(b) with appropriate integral limits.

## 3. Polarizations, Strains and Energy Predictions in $LiNbO_3$ and $LiTaO_3$ domain walls

Using the material constants listed in Table 2, the variation of the free energy, polarization, and strains as a function of domain wall orientation was calculated for both $LiNbO_3$ and $LiTaO_3$. These results are presented and discussed below.

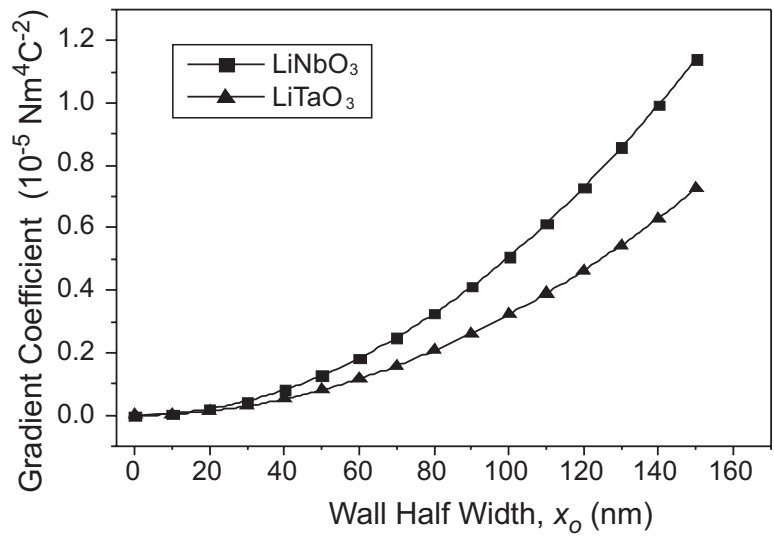

**Figure 4: Gradient coefficient, $g_1$, as a function of wall width, $x_o$**

Figure 4 shows a plot of the gradient term, $g_1$ as a function of the wall width, $x_o$. The domain wall width, the distance over which the polarization reverses, has been measured by Bursill et al. to have an upper limit of 0.28 nm using high-resolution TEM images in lithium tantalate (isomorphous to lithium niobate).[17] Taking this as the wall width, $2x_o$, for both materials, the upper limit for the gradient term is estimated as $3.98 \times 10^{-11}$ $Nm^4/C^2$ and $2.53 \times 10^{-11}$ $Nm^4/C^2$ for $LiNbO_3$ and $LiTaO_3$, respectively.

Since the theory does not include any energy contributions from non-stoichiometry related defects, *a direct comparison of the properties calculated below can be made only with the stoichiometric compositions of these materials*.

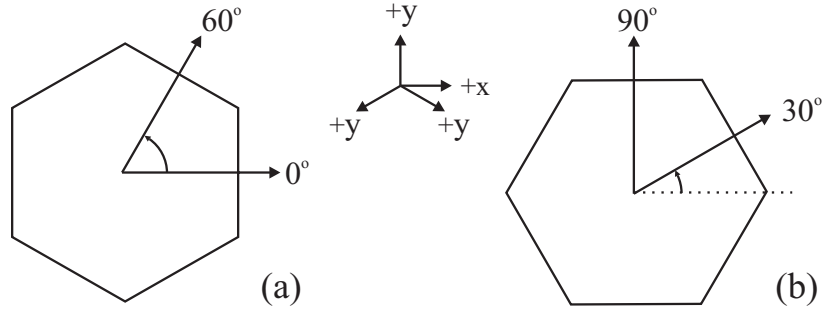

**Figure 5: Hexagonal wall orientations with wall normals for (a) *y*-walls and (b) *x*-walls.**

Two types of walls are of special interest in these materials: The six "*y*-walls" lying in the crystallographic *y-z* planes with wall normals at $\theta = m\pi/3$ as shown in Figure 5(a), and the six "*x*-walls" lying in the crystallographic *x-z* planes with wall normals at $\theta = (\pi/6 + m\pi/3)$ as shown in Figure 5(b), where *m* is an integer from 0-5. The stoichiometric crystals of both LiNbO$_3$ and LiTaO$_3$ possess domain orientations with *y*-walls. It is important to note for the rest of this paper that the angular dependence always refers to the orientation of the *normal* to the domain wall within the *x-y* plane.

**3.1 Polarizations**

Figure 6 shows the spontaneous polarization *P* as a function of normalized distance, $x_n/x_o$, perpendicular to a domain wall according to Eq. (31a). This variation is the same for all orientations $\theta$ of the domain wall in the *x-y* plane. The saturation polarization $P_s$ far from the domain wall is ±0.5 C/m$^2$ for LiTaO$_3$[14] and 0.75 C/m$^2$ for LiNbO$_3$.[14] The corresponding plot for Eq. (32) is a square-wave pattern with alternating kink anti-kink like profiles. An anti-kink is just the negative profile of Figure 6.

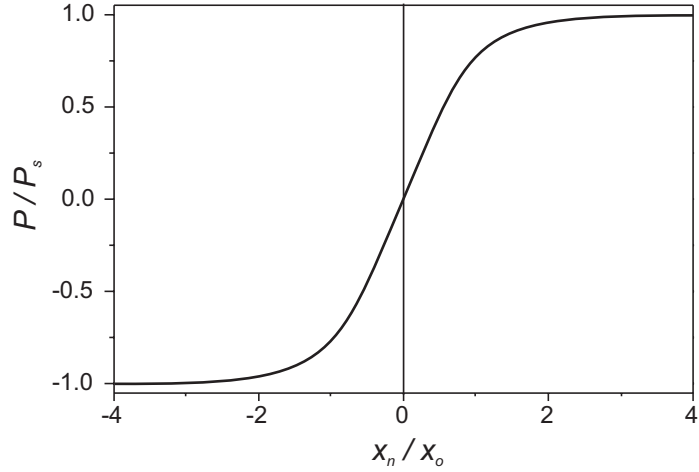

**Figure 6: The variation of the normalized polarization, $P/P_s=tanh(x_n/x_o)$, across a single 180° ferroelectric domain wall.**

In the absence of a domain wall, the polarizations in the *x-y* plane, $P_n$ and $P_t$, do not exist. However, they are non-zero in the vicinity of a domain wall, and disappear away from the wall. The magnitude and direction of these polarizations are dependent on the normal to the wall orientation, $\theta$. This is shown in a quiver plot in Figure 7(a) (for LiTaO$_3$) and Figure 7(b) (for LiNbO$_3$), where the in-plane polarization $P_{in-plane} = \sqrt{P_n^2 + P_t^2}$ is plotted as arrows. The length and direction of the arrows, respectively, represent the magnitude and direction of the vector $\vec{P}_{in-plane}$ in the *x-y* plane. The circle in the plot represents the schematic of a hypothetical circular domain wall. Figure 7(c) is a polar plot of the maximum amplitude of $\vec{P}_{in-plane}(\theta)$ for LiNbO$_3$ and LiTaO$_3$.

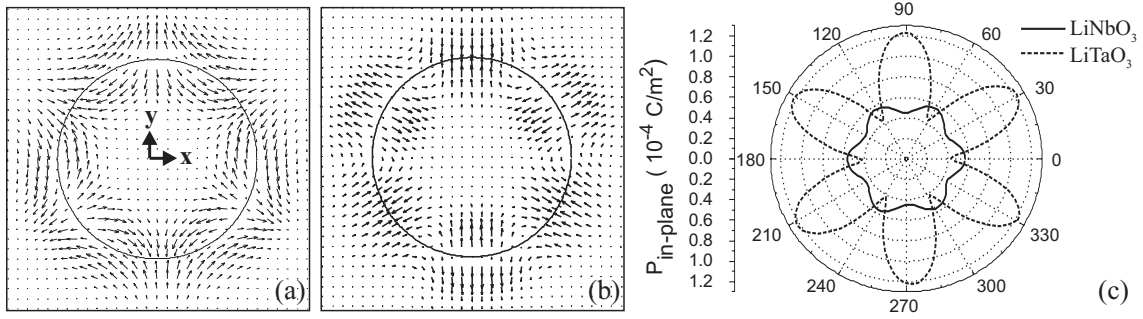

**Figure 7:** In-plane polarizations, $P_{in\text{-}plane}$, for (a) LiTaO$_3$ and (b) LiNbO$_3$. (c) Shows the maximum magnitude of the in-plane polarization for LiNbO$_3$ and LiTaO$_3$.

It is seen in Figure 8(a) that the *x*-walls have $\vec{P}_{in-plane} = \vec{P}_n$ and in Figure 8(b) the *y*-walls have $\vec{P}_{in-plane} = \vec{P}_t$. This is shown in Figure 8 for LiTaO$_3$ but is also true for LiNbO$_3$. In addition, these in-plane polarizations can also form *in-plane* antiparallel domain walls in the *x-y* plane. The $\vec{P}_n$ and $\vec{P}_t$ vectors reverse directions on crossing such a domain wall along the $x_n$ direction. The variation of $\vec{P}_n$ and $\vec{P}_t$ as a function of $x_n$ is shown in Figure 9(a) and in Figure 9(b), respectively for LiTaO$_3$. Again, Figure 9 is valid for LiNbO$_3$ as well by changing the sign and magnitudes of $\vec{P}_n$ and $\vec{P}_t$ for each of the curves in accordance with Figure 7(c).

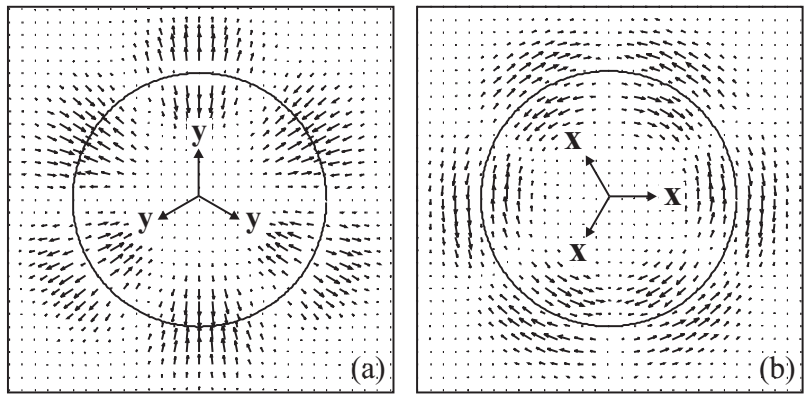

**Figure 8: (a) The normal polarizations, $P_n$, and (b) The transverse polarizations, $P_t$, for LiTaO$_3$. LiNbO$_3$ shows a similar symmetry but with the orientation of the vectors reversed.**

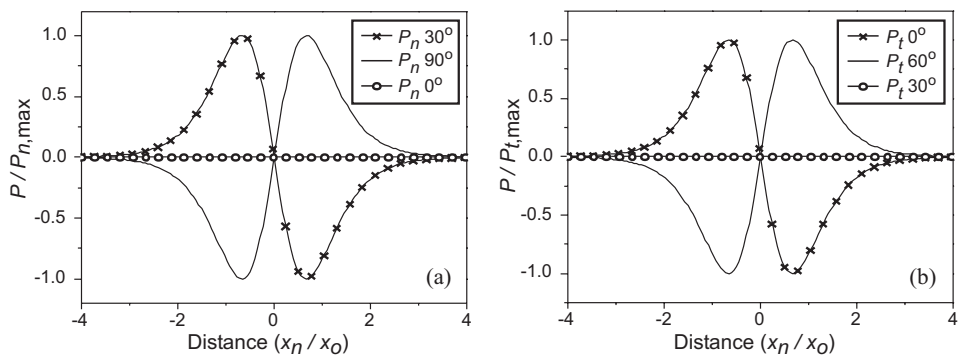

**Figure 9: Normalized in-plane polarizations as a function of $x_n$ in LiTaO$_3$. (a) Plot of normal polarizations, $P_n$, for different angles $\theta$. (b) Plot of transverse polarizations, $P_t$, for different angles $\theta$. Every 5th point is marked.**

A significant feature of these plots is that the in-plane polarizations at the *x*-walls are perpendicular to the wall, and are oriented in a head-to-head or tail-to-tail configuration across the walls. These domain walls must therefore be *electrically charged walls*. On the other hand, the *y*-walls have in-plane polarizations that are parallel to these walls, thereby creating *uncharged* walls. Therefore, the *x*-walls must have additional electrostatic wall energy as compared to *y*-walls; the energy arising from the divergence of in-plane polarization at the wall. This is a significant feature that is further discussed in Section 3.3.

**3.2 Strains**

In the absence of a domain wall (the homogeneous case), the spontaneous strains in LiNbO$_3$ and LiTaO$_3$ are (1) an isotropic strain $\varepsilon_n=\varepsilon_t=\lambda_1$ in the crystallographic *x-y* plane [see Eq. (13)], and (2) a normal strain $\varepsilon_3=\lambda_2$ in the *z*-direction [see Eq. (14)]. No shear strains exist [Eq. (9)].

In the presence of a single infinite domain wall, the strains in the domain wall region are different from the homogeneous strains far away from this wall. Since the domain wall plane *t-z* is considered infinite in both the *t* and the *z*-coordinate axes, the strains $\varepsilon_t$ and $\varepsilon_3$ and the shear strain $\tilde{\varepsilon}_4$ in the *t-z* plane of the domain wall do not change from their homogeneous values [see Eqs. (13)-(14) and (16)]. However, the strain $\varepsilon_n$ (strain normal to the domain wall in the direction $x_n$), shear strain in the *n-z* plane, $\tilde{\varepsilon}_5$, and shear strain in the *n-t* plane, $\tilde{\varepsilon}_6$, change from their homogeneous values by amounts given by Eq. (25).

The change in the normal strain, $\Delta\varepsilon_n(\theta)$, for both LiNbO$_3$ and LiTaO$_3$ is shown in Figure 10(a) and (b), respectively. The strains $\tilde{\varepsilon}_5(\theta)$ and $\tilde{\varepsilon}_6(\theta)$ at the center of the wall ($x_n=0$) are shown as polar plots in Figure 11(a) and (b), respectively.

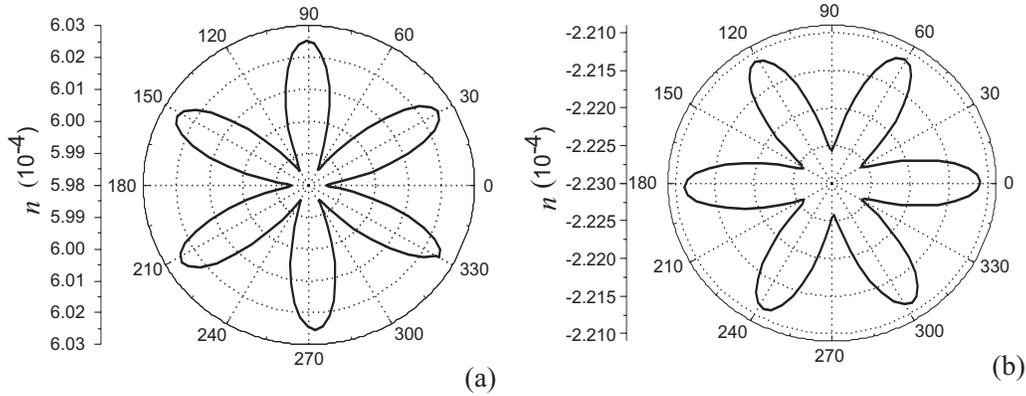

**Figure 10:** Change in the normal strain, $\Delta\varepsilon_n$, at the wall ($x_n = 0$) for (a) LiNbO$_3$ and (b) LiTaO$_3$.

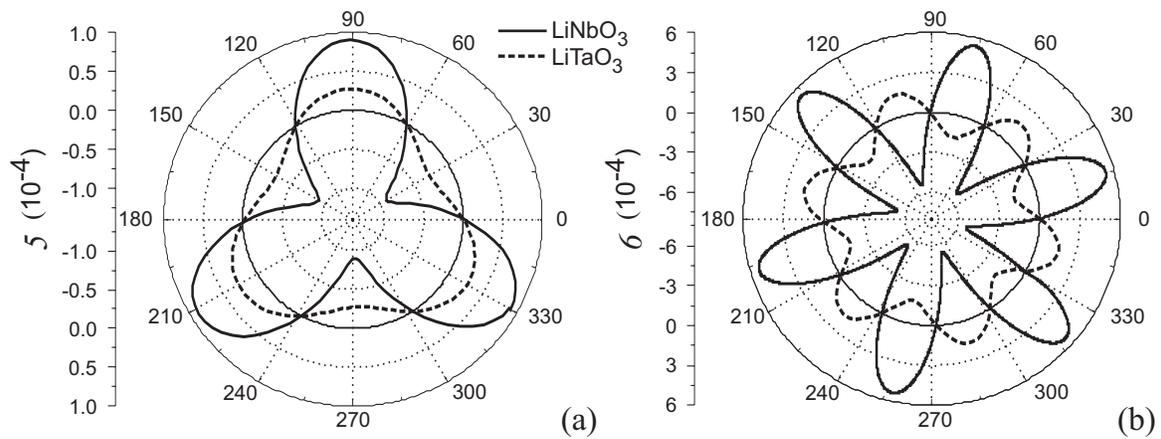

**Figure 11:** Strains at the wall ($x_n = 0$) for (a) $\tilde{\varepsilon}_5$ and for (b) $\tilde{\varepsilon}_6$. Note the circle in both figures represents zero strain.

The variation of these strains as a function of the normalized coordinate $x_n/x_o$ perpendicular to the domain wall in LiTaO$_3$ is plotted in Figure 12(a) and (b) for *x*-walls and *y*-walls, respectively. The corresponding plots for LiNbO$_3$ are shown in Figure 13.

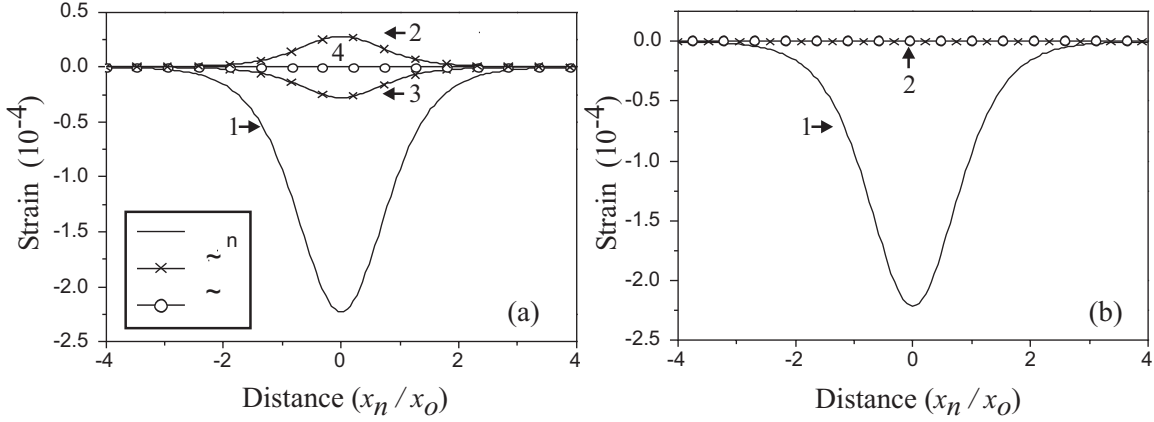

**Figure 12:** The strain in LiTaO$_3$ at (a) x-walls, where curve 1 is $\Delta\varepsilon_n$ for $\theta=30$ and 90°, curve 2 is $\tilde{\varepsilon}_5$ for $\theta=90°$, curve 3 is $\tilde{\varepsilon}_5$ for $\theta=30°$, and curve 4 is $\tilde{\varepsilon}_6$ for $\theta=30$ and 90°. The y-walls are shown in (b), where curve 1 is $\Delta\varepsilon_n$ for $\theta=0$ and 60°, and curve 2 is $\tilde{\varepsilon}_5$ and $\tilde{\varepsilon}_6$ for $\theta=0$ and 60°. Every 10th point is marked.

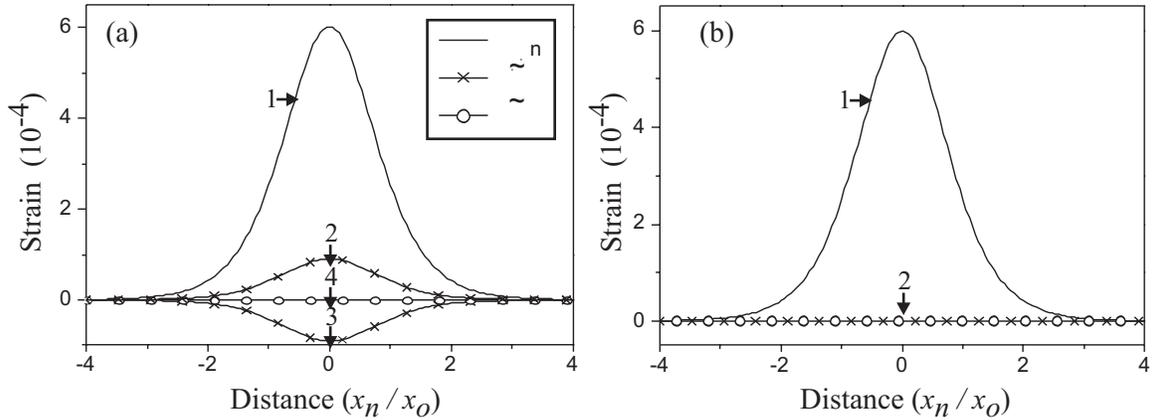

**Figure 13:** The strain in LiNbO$_3$ at (a) x-walls, where curve 1 is $\Delta\varepsilon_n$ for $\theta=30$ and 90°, curve 2 is $\tilde{\varepsilon}_5$ for $\theta=90°$, curve 3 is $\tilde{\varepsilon}_5$ for $\theta=30°$, and curve 4 is $\tilde{\varepsilon}_6$ for $\theta=30$ and 90°. The y-walls are shown in (b), where curve 1 is $\Delta\varepsilon_n$ for $\theta=0$ and 60°, and curve 2 is $\tilde{\varepsilon}_5$ and $\tilde{\varepsilon}_6$ for $\theta=0$ and 60°. Every 10th point is marked.

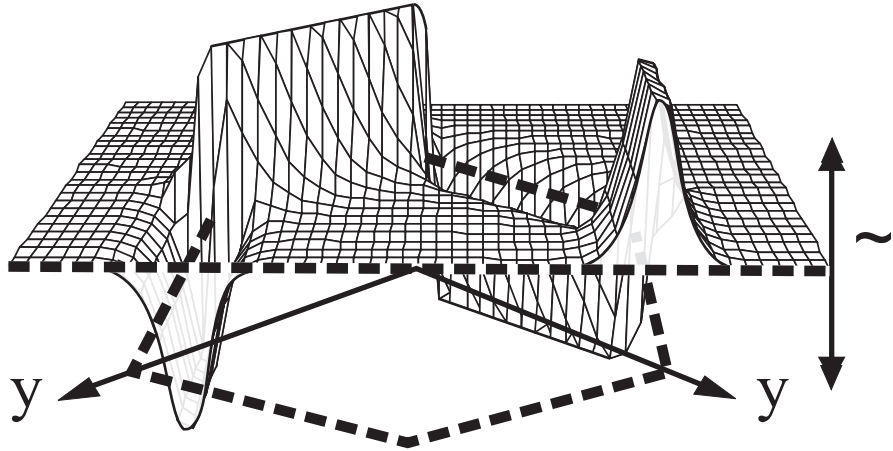

**Figure 14: Strain, $\widetilde{\varepsilon}_5$, for a theoretical *x*-wall shown as dotted lines in LiTaO$_3$. The horizontal dashed line is a cut through hexagon along the *x* direction. At the corners of the domain walls are high energy points as the sign of the strain switches.**

Some significant features are revealed in Figures 9-13 for both the *x*-walls and the *y*-walls.

(1) The shear strain $\widetilde{\varepsilon}_6 = 0$ in the *n-t* plane for the *x*-walls as well as the *y*-walls in both materials.

(2) The shear strain $\widetilde{\varepsilon}_5$ (shear strain in the *n-z* plane) is zero for the *y*-walls in both materials. However, this strain is non-zero for the *x*-walls. In addition, the sign of the shear strain $\widetilde{\varepsilon}_5$ changes from *positive* for the three *x*-walls at $\theta=(\pi/2+2m\pi/3)$, *m=0-2* to *negative* for the three *x*-walls at $\theta=(\pi/6+2m\pi/3)$, *m=0-2*. This is shown in Figure 11(a). Although the calculations are done for domain walls that are infinite in the lateral extent (along the *t*-axis), if we imagine the formation of a hexagonal domain by bringing together the six *x*-walls, every adjacent hexagonal face will have a different sign for the shear strain, $\tilde{\varepsilon}_5$ as shown in Figure 14. The above discussion is valid for both materials.

(3) The change in the normal strain $\Delta\varepsilon_n$ is *negative* for LiTaO$_3$ and *positive* for LiNbO$_3$ for all orientations of the domain wall. Since the homogeneous strain $\varepsilon_n$ in both materials is positive (net tensile strain; see Section 2.2), this indicates that the normal tensile strain $\varepsilon_n$ in the domain wall region is *lower* than the bulk value (by ~34% at the domain wall) in LiTaO$_3$ and *higher* in the domain wall region (by ~64% at the domain wall) in LiNbO$_3$ compared to the bulk value.

**3.3 Free Energy Anisotropy**

The free energy of the domain wall, $F_{DW}$, is numerically calculated from Eq. (6) in combination with Eq. (31b). This requires a knowledge of the gradient term $g_1$, which is not experimentally known, but was estimated earlier from the TEM measured atomic positions as $g_1 \leq 3.98 \times 10^{-11}$ Nm$^4$/C$^2$. For further calculations, we assume a value of $g_1 = 4 \times 10^{-11}$ Nm$^4$/C$^2$. While the absolute magnitude of free energy depends on the magnitude of the gradient term, the results discussed below deal with the energy anisotropy as a function of domain wall orientation angle, $\theta$. This energy anisotropy is characterized by the quantity $\Delta F_{DW} = [F_{DW}(\theta) - F_{DW}(0°)]$ which is calculated with respect to the minimum free energy which occurs at the y-walls. The symmetry of the dependence of $\Delta F_{DW}$ on the angle $\theta$ is found to be independent of the absolute value of the gradient term.

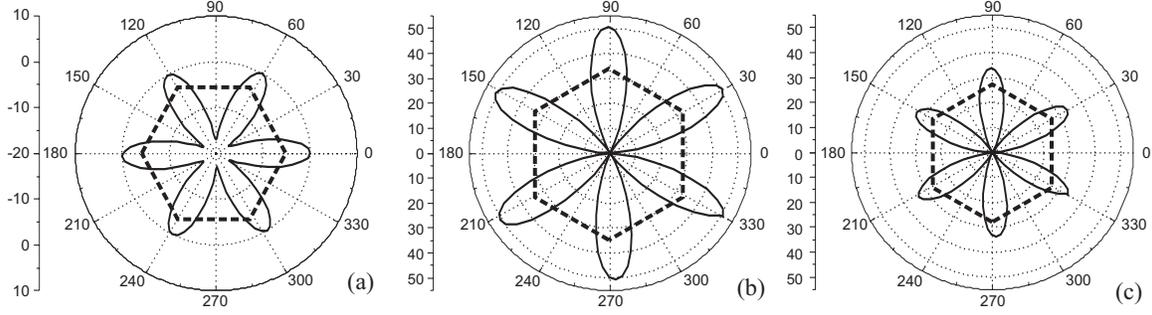

**Figure 15:** Energies of domain walls in LiTaO3 relative to 0°. (a) shows the normalized change in free energy, $\Delta F_{DW}$, (b) shows the depolarization energy, $\Delta F_d$ and (c) is the normalized change in the total energy, $\Delta F_{total} = \Delta F_{DW} + \Delta F_d$. Note that (b) and (c) have the same scale, while (a) does not. Units in all plots are J/m³. The dotted hexagon represents the low energy domain wall configuration for each plot.

Figure 15(a) and Figure 16(a) show a polar plot of the free energy $\Delta F_{DW}$ calculated by combining Eqs. (6) and (31b), as a function of domain wall normal orientation, $\theta$, with respect to the crystallographic $x$ axis for LiTaO$_3$ and LiNbO$_3$, respectively. The variation of domain wall energy is $\Delta F_{DW}/F_{mean} \sim 10^{-7}$, where $F_{mean} = \langle F_{DW}(\theta) \rangle$. Though small in magnitude, it was confirmed that the angular variation of $\Delta F_{DW}$ shown in Figure 15 is *not* a numerical artifact, since the polar symmetry of the energy plot was found to be insensitive to large variations in input parameters. Changing each of the physical constants individually in Table 2 did not change the symmetry of $\Delta F_{DW}/F_{mean}$. For example, changing the coefficients $c_{11}$, $c_{13}$, $c_{33}$, $c_{14}$, $q_{33}$, $q_{42}$, $q_{44}$, or $e_{11}$ by a factor 0.01 to 30 slightly changed the magnitude but did not change the symmetry of the free energy. The free energy was more sensitive to the coefficients $c_{12}$, $c_{13}$, $q_{31}$, and $e_{33}$, with the symmetry changing only if the coefficients were multiplied by a factor less than 0.6 or greater than 1.5. However, the changes in the physical constants needed to induce symmetry changes are very large and unphysical. Further, our calculations have a higher precision than the observed variation – the numerical variation is $\sim 10^2$ while calculations are carried out to $10^{-16}$. These results give us confidence in the energy anisotropy plots shown in Figures 15 and 16.

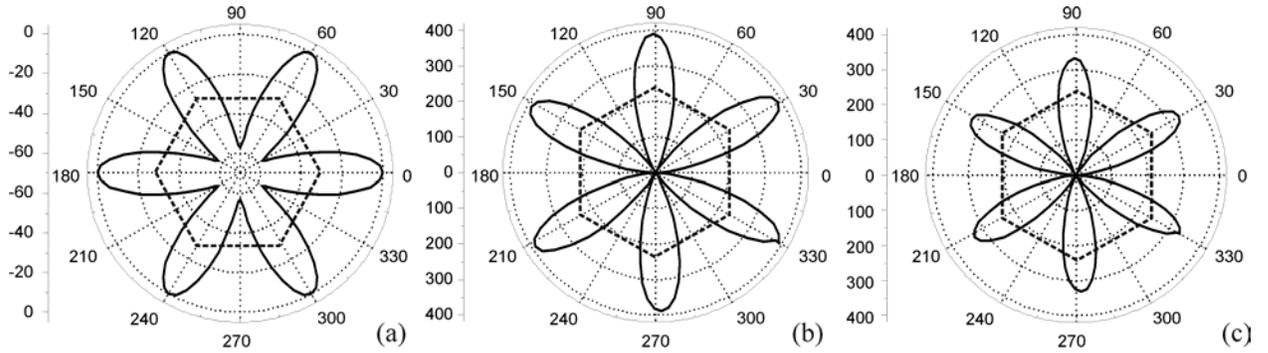

**Figure 16: Energies of domain walls in LiNbO$_3$ relative to 0º. (a) Shows the normalized change in free energy, $\Delta F_{DW}$, (b) shows the depolarization energy, $\Delta F_D$ and (c) is the normalized change in the total energy, $\Delta F_{total} = \Delta F_{DW} + \Delta F_d$. Note that (b) and (c) have the same scale, while (a) does not. Units in all plots are J/m$^3$. The dotted hexagon represents the low energy domain wall configuration for each plot.**

The change in free energy, given in Figure 15(a) and Figure 16(a) for LiNbO$_3$ and LiTaO$_3$, exhibits a six-fold symmetry with six energy minima at $\theta = (\pi/6 + m\pi/3)$ where $m=0$ to 5. These orientations correspond to $x$-walls, domain walls in the crystallographic $x$-$z$ planes with the wall normal in the $\pm y$ directions. Note that the six-fold symmetry of the lobes preserves the mirror symmetry about the three crystallographic $y$-axes. We note that the six-sided hexagonal domain that can be formed with these six minimum energy domain wall configurations *does not* correspond to the actual domain wall shapes observed experimentally in stoichiometric LiNbO$_3$ and LiTaO$_3$, as shown in Fig. 2.

One of the energy contributions missing in Eq. (6) is the depolarization energy at a domain wall introduced by the variation of in-plane polarization $P_n$ across the domain wall in the direction $x_n$. In other words, an additional depolarization energy term proportional to $P_{n,n}^2$, which was originally ignored, needs to be accounted for. This energy as a function of distance normal to the domain wall, $x_n$, is calculated starting from Gauss's law given as

$$E(x_n) = -\frac{P_n(x_n)}{\varepsilon_0}. \tag{33}$$

where $E(x_n)$ is the depolarizing field arising from the polarization, $P_n(x_n)$.[18] For a one dimensional case, where the electric field and polarization are zero at $\pm\infty$ for the normal components of electric field and polarization, the energy per unit area for a wall slice of $dx_n$ at $x_n$ is

$$\frac{\varepsilon_o}{2} E^2(x_n)\, dx_n = \frac{P_n^2(x_n)}{2\varepsilon_0}\, dx_n. \tag{34}$$

The depolarization energy per unit volume of the entire wall region is given by

$$F_d = \frac{1}{\Delta x} \int_{-\Delta x/2}^{\Delta x/2} \frac{P_n^2(x_n)}{2\varepsilon_0}\, dx_n \tag{35}$$

which is the depolarization energy per unit volume in units of J/m$^3$. The integration window, $\Delta x = 4x_o$, was chosen as in Eq. (31b). The depolarization energy in Eq. (35) is calculated numerically from the normal polarization, $P_n$, shown in Figure 9 as a function of distance, $x_n$, from the wall.

Figure 15(b) and Figure 16(b) show the depolarization energy, $\Delta F_d = F_d(\theta) - F_d(0°)$. It can be seen from these plots that the minimum energy is rotated 60° from the minimum energy configuration shown in Figure 15(a) and Figure 16(a). The depolarization energy favors $y$-domain walls in the crystallographic $y$-$z$ planes with the wall normal in the $\pm x$ directions. Since the change in the depolarization energy is larger than the change in the domain wall free energy, the resulting total energy, $\Delta F_{total} = \Delta F_{DW} + \Delta F_d$, have symmetry that favors $y$-walls as shown in Figure 15(c) and Figure 16(c).

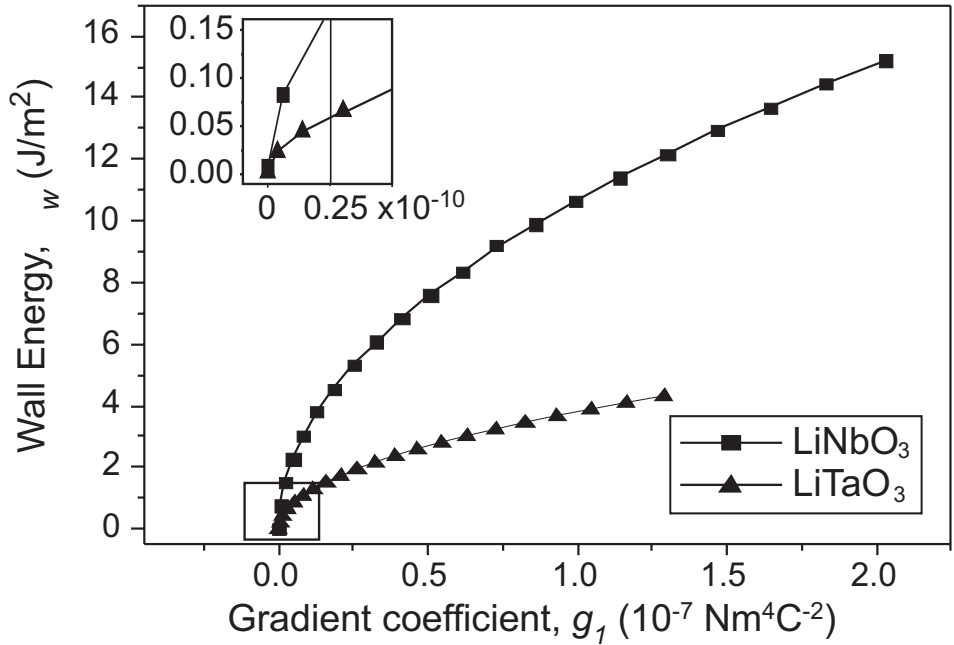

**Figure 17: Domain wall energy per unit , $F_{DW}$, as a function of the gradient coefficient $g_1$. The inset of the figure is an expansion of the plot near zero and the vertical line is the upper estimate of $g_1$ calculated from the domain wall width from the literature.** [17]

Figure 17 shows the plot of total free energy as a function of the gradient coefficient $g_1$. Using the upper limit on the width of the domain wall as 0.28 nm in LiTaO$_3$[17], the gradient energy is 2.53x10$^{-11}$ Nm$^4$/C$^2$. Using this value, the calculated domain wall energy, $F_{DW}$, in LiTaO$_3$ is ~60 mJ/m$^2$. Experimental estimates of domain wall energy vary. Using the activation field for the experimentally measured exponential dependence of sideways domain velocity in an applied electric field in *congruent* LiTaO$_3$, and following the Miller-Weinreich theory,[19] Gopalan et al. have estimated the wall energy to be ~35 mJ/m$^2$.[20] Following this analysis and using data for the wall velocity in stoichiometric crystals[4], the wall energy in *stoichiometric* composition crystals (which is the correct material composition for comparison with the presented calculations) is calculated as ~9 mJ/m$^2$. This estimate considers only the polarization and depolarization energies, and ignores strain, coupling and gradient energies. On the other hand, using the curvature of a pinned domain wall under an external field in congruent LiTaO$_3$, and modeling the process as a tradeoff between a decrease in polarization energy and an increase in domain wall energy, Yang and Mohideen estimated the wall energy as $F_{DW}$ ~200-400 mJ/m$^2$.[21] Yet another estimate based on optical birefringence at the domain wall over a 3 μm width in congruent LiTaO$_3$ yields an electrostatic energy of ~240 mJ/m$^2$.[6] The estimation of wall energy in this study is near the lower end of experimental estimations.

## 4. Discussion

Although differences between lithium niobate and lithium tantalate in the preceding analysis are slight, we find it important to highlight the major differences.

With respect to polarizations, each material shares the same symmetry, with charged domains walls for *x*-wall orientations and uncharged walls for *y*-wall orientations. However, in addition to differences in the magnitudes of the polarization, the sign of each is different with polarization in head-to-head configuration in LiTaO$_3$ with domain wall normals at 30º, 150º, and 270º, and in LiNbO$_3$ at 90º, 210, and 330º. The change in the normal strain, $\Delta\varepsilon_n$ is negative for LiTaO$_3$ and positive for LiNbO$_3$ for all orientations. This normal strain is lowest for *x*-wall orientations in LiTaO$_3$ and *y*-wall orientations in LiNbO$_3$.

By combining information from the polarizations, strains, and energies of the domain walls as functions of wall angle, comments can be made on preferred orientations. Considering only the free energy contribution as in Eq. (6), it is found that the minimum energy configuration is for the *x*-wall orientations as shown in Figure 15(a) and Figure 16(a). However, the *x*-walls are charged domain walls due to head-to-head or tail-to-tail in-plane polarization configurations on crossing the domain wall. This *in-plane* polarization leads to high depolarizing energy for the *x*-walls, giving the total energy of the domain walls a minimum for *y*-wall orientations for both LiTaO$_3$ and LiNbO$_3$ as shown in Figure 15(c) and Figure 16(c).

Considering the strain, either domain wall orientation (*x*-wall or *y*-wall) has a zero $\widetilde{\varepsilon}_6$ component, which is strain in the $x_t$-$x_n$ plane. However, the $\widetilde{\varepsilon}_5$ strain, strain in the $x_n$-$z$ plane, is non-zero for the *x*-walls and contrary in sign for adjacent hexagonal faces and is as pictured in Figure 14. The vertices of a hexagon formed by these *x*-walls would therefore be high-energy points, requiring a screw-like defect at that site to accommodate the change in the sign of this shear strain. On the other hand, there are no such restrictions at the vertices of a hexagonal domain formed by the *y*-walls and lower energy vertices result.

The free energy and strain analysis of the crystallographic contributions therefore supports the physical reality of *y*-walls being preferred over *x*-walls in *stoichiometric* crystals of both lithium niobate and lithium tantalate.

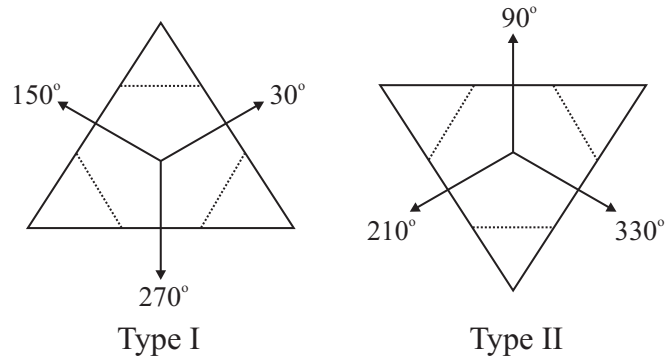

Type I　　　　　　　　　　Type II

**Figure 18: Two possible sets of triangular *x*-walls. The dotted walls in each case outline the hexagonal *x*-wall configuration for clarity.**

This analysis, however, ignores non-stoichiometric defect complexes present in the crystal structure.[4] These defects drastically change the poling kinetics, and in the case of lithium tantalate, also change the preferred domain wall orientation. In this case, instead of hexagonal *y*-wall domain shapes seen in the stoichiometric crystals, triangular *x*-walls are preferred in *congruent* composition of $Li_{0.95}Ta_{1.01}O_3$, as shown in Figure 2. It is clear that these defects in combination with the previously highlighted differences between the crystals, favor formation of triangular domains formed by one of two sets of *x*-walls, as shown in Figure 18. However, neglecting for the moment, both the non-stoichiometric defects and the interactions of domain walls, it is interesting to think about the *x*-wall orientations. Domains with *x*-wall orientations have in-plane polarization normal to the domain wall and non-zero strain, $\tilde{\varepsilon}_5$, in the $x_n$-$z$ plane. Since the sign of this strain is contrary on adjacent faces (as in Figure 14), triangles composed of every other domain wall orientation have the same sign of stress on all adjacent faces eliminating the high strain points at the corners of a hexagon formed by *x*-walls.

Therefore, one can conclude that within this theoretical framework, if the *x*-walls are preferred at all, they should occur as triangles, unless there are screw-like dislocations at the vertices of a hexagon to facilitate a hexagonal domain composed of *x*-walls. Nevertheless, the two sets of *x*-walls are degenerate in energy within the free energy described, and therefore they might be expected to occur with equal probability. However, in congruent crystals, one of these sets (Type I) is clearly preferred over the other (Type II). The presence of non-stoichiometric defects therefore appears to prefer one set over the other. In order to understand this preference, one will have to better understand the nature of these organized point defects, and their contribution to the free energy, which is expected to be anisotropic as well. We note that many symmetry allowed higher order gradient energy terms exist and in the preceding analysis, we have only considered the lowest order energy terms [Eq. (5)]. Two such higher order terms with the proper 3-fold and 6-fold symmetries are given as

$$F_{G:3fold}(P_{i,j}) = g_2\left(P_{z,3}^3\right) + g_3\left(6P_{z,x}^2 P_{z,y} - 2P_{z,y}^3\right) \tag{36}$$

$$F_{G:6fold}(P_{i,j}) = g_2\left(P_{z,3}^6\right) + g_6\left(2P_{z,y}^6 - 2P_{z,x}^6 + 30P_{z,x}^4 P_{z,y}^2 - 30P_{z,x}^2 P_{z,y}^4\right) \tag{37}$$

where $g_3$ ad $g_6$ are the 3-fold and 6-fold gradient coefficients. It was noted that these terms, when included in time-dependent Ginzburg-Landau (TDGL) simulations, can result in the evolution of hexagonal or trigonal domain shapes.[22] If the 3-fold term dominates, triangular domains evolve. Similarly, hexagonal domains evolve if the 6-fold energy term dominates. One possibility is that the non-stoichiometric point defects influence these higher order energy terms to give rise to symmetries not obvious in the one-dimensional analysis presented in this paper. The presented model is valuable, however, in understanding the intrinsic structure of a domain wall expected without the presence of extrinsic defects, external fields, or higher order energy terms whose coefficients are not known experimentally.

The importance of the defects to the observed domain features is further supported by the temperature effects observed in the LiTaO$_3$ system, where at higher temperatures the *y* walls are favored orientation as shown in Figure 2(c),(d). This indicates the change in domain shape could be due to the decrease or disappearance of the influence of the defect dipoles. One of the proposed models for the defect complex is Nb or Ta antisites $\left(Nb_{Li}^{4+} \text{ or } Ta_{Li}^{4+}\right)$ surrounded by lithium vacancies $\left(V_{Li}^{-}\right)$ with a charge balance of $4[Nb_{Li}^{4+}] = [V_{Li}^{-}]$.[4] At temperatures above 150ºC, the internal field related to these lithium vacancies has been shown to quickly reorient and assume a low energy configuration.[23] At room temperature (25ºC) however, these defects are frozen and form aggregated defect dipole complexes. One of the clear correlations therefore is that changes in domain shapes in congruent LiTaO$_3$ with temperature (as shown in Figure 2) are accompanied by changes in defect complexes with temperature.

## 5. Conclusions

The preferred domain wall shapes of ferroelectrics LiNbO$_3$ and LiTaO$_3$ have been analyzed by taking into account the free energy of the system. A theoretical framework has been developed to analyze the polarizations, strains, and energies associated with a domain wall of arbitrary orientation in both lithium niobate and lithium tantalate. It was found that *x*-walls are charged domain walls due to head-to-head or tail-to-tail in-plane polarizations, maximum strains, and maximum total free energy. In contrast, the *y*-walls show a minimum in strains, zero head-to-head or tail-to-tail in-plane polarization, and a minimum in the total free energy. The *y*-walls are therefore the preferred orientations in stoichiometric compositions and this is supported by experimental observations of such hexagonal domains composed of *y*-walls in the stoichiometric compositions of these materials. This analysis does not directly consider the interaction of multiple domain walls as well as the influence of non-stoichiometric point defects present in the congruent compositions of these materials. These point defects have been proposed to be organized into defect complexes,[4] and probably have different defect symmetries that lead to triangular domains in congruent lithium tantalate. It was found that domains created at temperatures higher than 125ºC in $LiTaO_3$ formed domains composed of *y*-walls favored by the stoichiometric crystals, instead of forming *x*-walls normally seen when created at 25ºC. This indicates that the nature of the influence of the defects on the wall orientations is changing with temperature. The exact mechanism of defect-domain wall interactions is presently unknown in these materials and will require understanding the structure and symmetry of defects themselves on the atomic scale.

## Acknowledgements

Scrymgeour and Gopalan would like to acknowledge support from grant numbers DMR-9984691, DMR-0103354, and DMR-0349632. This work was supported in part by the U.S. Department of Energy.

# Appendix A

The constants $v_i$ and $\mu_{ij}$ used in Eq, (23) are

$v_1 = -\gamma_3 m_{31} \cos(3\theta) - \gamma_3 m_{11} \sin(3\theta) - \gamma_4 m_{21}$

$v_2 = -\gamma_3 m_{11} \cos(3\theta) + \gamma_3 m_{31} \sin(3\theta)$

$\mu_{11} = \gamma_3 m_{32} \cos(3\theta) + \gamma_3 m_{12} \sin(3\theta) + \gamma_4 m_{22}$

$\mu_{12} = \gamma_3 m_{33} \cos(3\theta) + \gamma_3 m_{13} \sin(3\theta) + \gamma_4 m_{23}$

$\mu_{21} = \gamma_3 m_{12} \cos(3\theta) - \gamma_3 m_{32} \sin(3\theta)$

$\mu_{22} = \gamma_3 m_{13} \cos(3\theta) - \gamma_3 m_{33} \sin(3\theta)$

………………………………………………………………………………..(A.1)

The matrices $[\rho_{ij}]$ and $[\phi_{ij}]$ in Eqs. (24) and (25) are listed below.

$\rho_{11} = -\dfrac{v_1 P_h^2}{\alpha_3}$

$\rho_{12} = \dfrac{v_1}{\alpha_3} - \dfrac{(v_1\mu_{22} - v_2\mu_{12})P_h^2}{\alpha_3^2}$

$\rho_{13} = \dfrac{v_1\mu_{22} - v_2\mu_{12}}{\alpha_3^2}$

$\rho_{21} = -\dfrac{v_2 P_h^2}{\alpha_3}$

$\rho_{22} = \dfrac{v_2}{\alpha_3} + \dfrac{(v_1\mu_{21} - v_2\mu_{11})P_h^2}{\alpha_3^2}$

$\rho_{23} = \dfrac{v_2\mu_{11} - v_1\mu_{21}}{\alpha_3^2}$

$\phi_{i1} = -m_{i1} P_h^2,$ \hspace{2cm} (i=1,2,3)

$$\phi_{i2} = m_{i1} + m_{i2}\rho_{11} + m_{i3}\rho_{21}, \qquad (i=1,2,3)$$

$$\phi_{i3} = m_{i2}\rho_{12} + m_{i3}\rho_{22}, \qquad (i=1,2,3)$$

$$\phi_{i4} = m_{i2}\rho_{13} + m_{i3}\rho_{23}, \qquad (i=1,2,3)$$

……………………………………………………………………………...(A.2)

[1] D.F. Morgan and D. Craig, in *Properties of lithium niobate* (INSPEC, IEE, 2002), pp. 243; D. Psaltis and G.W. Burr, Computer **31** (2), 52 (1998).

[2] Martin M. Fejer, G. A. Magel, Dieter H. Jundt, Robert L. Byer, IEEE Journal of Quantum Electronics **28** (11), 2631 (1992); G. Rosenman, A. Skliar, A. Arie, Ferroelectrics Review **1**, 263 (1999).

[3] Qibiao Chen, Yi Chiu, D.N. Lambeth, T. E. Schlesinger, D. D. Stancil, Journal of Lightwave Technology **12** (8), 1401 (1994); D. A. Scrymgeour, Y. Barad, V. Gopalan, K. T. Gahagan, Q. Jia, T. E. Mitchell; J. M. Robinson, Applied Optics **40** (34), 6236 (2001).

[4] Sungwon Kim, V. Gopalan, K. Kitamura, Y. Furukawa, Journal of Applied Physics **90** (6), 2949 (2001).

[5] Jung Hoon Ro, Tae-hoon Kim, Ji-hyun Ro, Myoungsik Cha, Journal of the Korean Physical Society **40** (3), 488 (2002).

[6] T.J. Yang, U. Mohideen, V. Gopalan, P.J. Swart, Physical Review Letters **82** (20), 4106 (1999).

[7] T. Jach, S. Kim, V. Gopalan, S. Durbin, D. Bright, Physical Review B **69** (6), 064113 (2004).

[8] A. Gruverman, O. Kolosov, J. Hatano, K. Takahashi, H. Tokumoto, Journal of Vacuum Science & Technology B **13** (3), 1095 (1995).

[9] A.F. Devonshire, Philosophical Magazine **42** (333), 1065 (1951).

[10] A.F. Devonshire, British Electrical and Allied Industries Research Association - Technical Reports, 24 (1951); Ennio Fatuzzo and Walter J. Merz, *Ferroelectricity*. (North-Holland Pub. Co., Amsterdam, 1967).

[11] H. L. Hu and L. Q. Chen, Materials Science & Engineering A **238**, 182 (1997).

[12] T. Yamada, Landolt-Bornstein. Numerical Data and functional relationships in science and technology, 149 (1981); J. Kushibiki, I. Takanaga, M. Arakawa et al., IEEE Transactions on Ultrasonics, Ferroelectrics and Frequency Control **46** (5),

1315 (1999); I. Takanaga and J. Kushibiki, IEEE Transactions on Ultrasonics Ferroelectrics and Frequency Control **49** (7), 893 (2002).
13  K. Kitamura, Y. Furukawa, K. Niwa, V. Gopalan, T. E. Mitchell, Applied Physics Letters **73** (21), 3073 (1998).
14  V. Gopalan, T.E. Mitchell, K. Kitamura, Y. Furukawa, Applied Physics Letters **72** (16), 1981 (1998).
15  Pei Chi Chou and Nicholas J. Pagano, *Elasticity : tensor, dyadic, and engineering approaches*. (Van Nostrand, Princeton, N.J., 1967).
16  M. Abramowitz and I. A. Stegun, in *Handbook of Mathematical Functions with Formulas, Graphs, and Mathematical Tables* (U.S. GPO, Washington, D.C., 1964).
17  L.A. Bursill and Peng Ju Lin, Ferroelectrics **70** (3-4), 191 (1986).
18  Charles Kittel, *Introduction to solid state physics*, 7th ed. (Wiley, New York, 1996).
19  R.C. Miller and G. Weinreich, Physical Review **117** (6), 1460 (1960).
20  V. Gopalan, T.E. Mitchell, K.E. Sickafus, Integrated Ferroelectrics **22** (1-4), 405 (1998).
21  T.J. Yang and U. Mohideen, Physics Letters A **250** (1-3), 205 (1998).
22  Sungwon Kim, PhD Thesis, Pennsylvania State University, 2003.
23  V. Gopalan and M.C. Gupta, Applied Physics Letters **68** (7), 888 (1996).